\documentclass[aps,prb,reprint,longbibliography,citeautoscript,
               floatfix,showkeys]{revtex4-2}
\usepackage[utf8]{inputenc}
\usepackage[T1]{fontenc}

\usepackage{graphicx}
\usepackage{amsmath}
\usepackage{siunitx}
\usepackage{booktabs}
\usepackage[normalem]{ulem}
\usepackage{hyperref}

\newcommand{\mytitle}{Triple Junctions as Dislocation-Like Defects:
  The Role of Grain Boundary Crystallography Revealed by Experiment
  and Atomistic Simulation}

\hypersetup{
  final,
  pdfborder={0 0 0},
  pdftitle={\mytitle},
  pdfauthor={Brink, Saood, Schweizer, Neugebauer, Dehm},
  colorlinks=true,
  urlcolor=blue,
  linkcolor=blue,
  citecolor=blue
}

\begin{document}
\frenchspacing
\raggedbottom

\title{\mytitle}

\author{Tobias Brink}
\email{t.brink@mpi-susmat.de}
\affiliation{Max Planck Institute for Sustainable Materials,
  Max-Planck-Stra\ss{}e 1, 40237 D\"usseldorf, Germany}

\author{Saba Saood}
\affiliation{Max Planck Institute for Sustainable Materials,
  Max-Planck-Stra\ss{}e 1, 40237 D\"usseldorf, Germany}

\author{Peter Schweizer}
\affiliation{Max Planck Institute for Sustainable Materials,
  Max-Planck-Stra\ss{}e 1, 40237 D\"usseldorf, Germany}

\author{Jörg Neugebauer}
\affiliation{Max Planck Institute for Sustainable Materials,
  Max-Planck-Stra\ss{}e 1, 40237 D\"usseldorf, Germany}

\author{Gerhard Dehm}
\email{dehm@mpi-susmat.de}
\affiliation{Max Planck Institute for Sustainable Materials,
  Max-Planck-Stra\ss{}e 1, 40237 D\"usseldorf, Germany}

\begin{abstract}
  Grain boundary networks and their evolution are strongly influenced
  by triple junctions. The defect nature of these line defects
  significantly affects the properties of the network, but they have
  not been fully characterized to date. Here, we use scanning
  transmission electron microscopy combined with atomistic computer
  simulations to investigate a triple junction at the atomic scale in
  an Al thin film with $\{111\}$ texture. Using sampling methods, we
  were able to construct the same junction structure as in the
  experiment within a computer model. We present a technique to
  calculate the Burgers vector of the triple junction. This allows us
  to connect the junction's dislocation character to the microscopic
  degrees of freedom of the joining grain boundaries. The junction
  line energy in the computer model can then be calculated using an
  embedded atom method potential. It follows the same laws as a bulk
  dislocation. Finally, we discovered a range of possible triple
  junctions for the observed grain boundaries, which vary in the
  magnitude of their Burgers vector. Interestingly, the experimentally
  observed junction is not the one with the smallest possible Burgers
  vector and energy. This suggests that the kinetics of transforming
  the junction line are likely too slow to be driven by the small
  energy contribution of the triple junction.
\end{abstract}

\keywords{
  grain boundary triple junction;
  degrees of freedom;
  Burgers vector;
  scanning transmission electron microscopy;
  molecular dynamics;
  grand-canonical sampling
}

\maketitle

\newcounter{supplfigctr}
\renewcommand{\thesupplfigctr}{S\arabic{supplfigctr}}
\newcounter{suppltabctr}
\renewcommand{\thesuppltabctr}{\mbox{S-\Roman{suppltabctr}}}
{\refstepcounter{supplfigctr}\label{fig:suppl:grip-S3}}
{\refstepcounter{supplfigctr}\label{fig:suppl:grip-S39}}
{\refstepcounter{supplfigctr}\label{fig:suppl:grip-S13}}
{\refstepcounter{supplfigctr}\label{fig:suppl:S3-shear}}
{\refstepcounter{supplfigctr}\label{fig:suppl:energy-dependence-on-center}}
{\refstepcounter{supplfigctr}\label{fig:suppl:relation-excess-to-line-energy}}
{\refstepcounter{supplfigctr}\label{fig:suppl:junction-search-thicknesses}}
{\refstepcounter{suppltabctr}\label{tab:suppl:csl-dsc}}
{\refstepcounter{suppltabctr}\label{tab:suppl:dof}}

\section{Introduction}

\vspace{-0.7\baselineskip}%
Grain boundaries (GBs) are common and important defects in
materials. Mechanical properties of metals, for example, are
influenced by the amount and type of GBs in the sample \cite{Hall1951,
  Petch1953, Randle2010, Krause2019}. GB networks evolve by the
formation of GBs during synthesis and later by their movement
\cite{Schuh2003b, Schuh2003a, Han2018}. However, GBs never end in
regions of defect-free bulk material, but end instead either at
surfaces or at so-called triple junctions, where three GBs meet in a
line \cite{Priester2013ch10}. Quadruple or higher junctions are
unstable \cite{Fortes1993} and the GB network in a material is thus
defined by the triple junctions \cite{Schuh2003b, Schuh2003a}. Indeed,
the properties of triple junctions in thin films, for example, might
strongly influence the overall GB network \cite{Patrick2023}. The
mobility of GBs is also affected by the drag of less mobile triple
junctions \cite{Czubayko1998, Gottstein1999, Gottstein2000}, giving
rise to the idea of GB junction engineering \cite{Gottstein2006}.

On the microscopic level, triple junctions represent objects that are
structurally and thermodynamically distinct from GBs
\cite{King1999}. Depending on the macroscopic and microscopic degrees
of freedom (DOFs) of the three joining GBs, the triple junction can
exhibit a dislocation character (translational DOFs) and/or a
disclination character (rotational DOFs) \cite{King1999,
  Shenderova1999, Shekhar2008}. This defect character can interact
with disconnections---responsible for GB mobility under stress
\cite{Han2018}---and is thereby of fundamental importance for
understanding GB network evolution from a mechanistic viewpoint
\cite{Thomas2019, Wei2020}.

Additionally, the junction retains its own DOFs, for example with
regards to its exact position, which also affect its core energy
\cite{King1999}. Indeed, the triple junctions must possess an energy
excess over the defect-free bulk material, but compared to the
abutting GBs, it can be more or less energetically favorable
\cite{King1999}. There is some discussion in the literature about the
latter point. It was found that atoms in triple junctions have
comparable energies to GB atoms \cite{Caro2001, Li2011b}. Others,
however, calculated or measured triple junction line energies and
obtained both positive \cite{Gottstein2010} and negative
\cite{Eich2016, Tuchinda2024} line energies. It is notable that often
only a single number is provided for the triple junction energy, but
if the triple junction has dislocation or disclination character, it
is connected to a strain field whose energy diverges with infinite
system size, like a dislocation \cite{HirthLothe1992}.

There is also a similarity between triple junctions on one hand, and
GB facet junctions \cite{Dimitrakopulos1997, Pond1997, Hamilton2003,
  Medlin2017, Hadian2018, Brink2024, Choi2025} or GB phase junctions
\cite{Frolov2021, Langenohl2022, Winter2025} on the other. All three
are discontinuities in the DOFs of the GB. Different GB phases
correspond to the thermodynamic states of a GB, characterized by
distinct microscopic DOFs, but may in some cases share the same
macroscopic DOFs \cite{Hart1968, Hart1972, Cahn1982, Rottman1988,
  Frolov2012, Frolov2012a, Frolov2021, Winter2025}. Facet junctions,
in contrast, necessarily represent a change of macroscopic DOFs
\cite{Dimitrakopulos1997, Pond1997, Hamilton2003, Medlin2017,
  Hadian2018, Brink2024, Choi2025}. The two types of ``double
junctions'' therefore also possess dislocation character and Burgers
vectors \cite{Dimitrakopulos1997, Pond1997, Hamilton2003, Medlin2017,
  Hadian2018, Brink2024, Choi2025, Frolov2021, Langenohl2022,
  Winter2025} and we will draw some analogies to triple junctions in
our analysis later.

In the present contribution, we investigate a specific triple junction
in Al, which has been recorded by atomic-resolution scanning
transmission electron microscopy (STEM). We thoroughly characterize
its Burgers vector content using atomistic computer simulations with
an empirical potential, which allows us to evaluate previous
assertions about the nature of the triple junction line energy. Our
findings for this concrete example case demonstrate methods of
simulating and analyzing the triple junction and put its defect
character in direct relation with the DOFs of the abutting GBs.

\section{Methods/Theory}

\vspace{-0.5\baselineskip}%
\subsection{Experimental}

Atomic resolution imaging of triple junctions requires precise
crystallographic and geometric alignment, such that all three grains
share a common zone axis (in our case $[11\overline{1}]$) and the
three tilt GBs meeting at the junction are oriented perfectly
edge-on. To satisfy these conditions, Al films with a pronounced
\{111\} texture were fabricated using the magnetron sputtering method,
with the deposition parameters described in Ref.~\cite{Saood2023}. The
crystallographic texture and GB types in the films were examined by
electron-backscattered diffraction (EBSD) at \SI{20}{kV} using a JEOL
JSM-6490 secondary electron microscope (SEM).

The triple junctions of interest were extracted using Xe focused ion
beam (FIB) milling performed on a Helios G3 Cx dual-beam SEM/FIB
system (Thermo Fisher Scientific). The milling parameters are listed
in Ref.~\cite{Saood2023}. The use of Xe ions avoids any Ga
implantation at GBs, which is known to compromise the investigation of
intrinsic triple junction structures in pure Al. Clean imaging of the
junction structure required considerable experimental selectivity. In
addition to the triple junction presented in the results section, two
additional junctions were also lifted out. However, in both cases, one
of the GBs was found to be locally deviated from the edge-on
condition, particularly in the vicinity of the junction. This
limitation occurred despite prior screening by EBSD.

The specimen was subsequently examined in STEM mode using a
probe-corrected FEI Titan Themis 80-300 (Thermo Fisher Scientific)
operated at \SI{300}{kV} with probe currents of
70--\SI{80}{pA}. High-angle annular dark-field (HAADF) images were
acquired using a Fishione detector with collection angles of
78--\SI{200}{mrad}. The images reported here were obtained by
averaging 50 to 100 consecutive frames.

\vspace{-0.5\baselineskip}%
\subsection{Simulation}

All simulations were performed with an embedded atom method (EAM)
potential for Al \cite{Zhakhovskii2009} as downloaded from the NIST
Interatomic Potentials Repository \cite{ipr}. This potential was found
to be the most accurate one to model GB phases in $\Sigma3$
$\langle111\rangle$ tilt boundaries \cite{Choi2025} and we therefore
also selected it for the present study.

In a first step, we indexed the experimental GBs and then ran a GB
structure search to find the matching GB phases to the experimental
observation. We used GRIP \cite{Chen2024} to search the structures,
which is a code that takes the macroscopic DOFs and constructs various
GB supercells (here we used variations from $1\times1$ to $5\times5$
in the GB plane). The boundary conditions were periodic along the
directions in the GB plane and we used open boundaries normal to the
GB. The code then samples various displacements and
insertions/deletions of GB atoms, runs a canonical MD simulation at
random temperature with 95\% probability, and finally minimizes the
atomic positions with regards to the potential energy. We then ordered
the resulting structures by their GB energy and found the
experimentally-observed structures among the low-energy results.

We tested the stability of the GB structures by running molecular
dynamics (MD) simulations at \SI{300}{K} using Nos\'e--Hoover
thermostats. The volume was scaled from \SI{0}{K} to the thermal
expansion of the bulk and then kept constant in the periodic
directions. We used a time integration step of \SI{2}{fs} for all
simulations.

For comparison with the experiment, we used STEM image simulations
with the abTEM code \cite{Madsen2021}. We simulated an electron probe
with a voltage of \SI{300}{keV}, a spherical aberration of
\SI{10}{\micro\meter}, a semiangle of \SI{18}{mrad}, a step size of
\SI{0.164}{\angstrom}, and an annular detector range from 90 to
\SI{150}{mrad}. Gaussian defocus and noise was applied after the image
simulations.

Afterwards, we assembled triple junctions using GBs with the same GB
phases and different microscopic DOFs (that are crystallographically
equivalent). The resulting sample is periodic along the shared tilt
axis direction $[11\overline{1}]$, initially having a thickness of one
unit cell. Then we cut a cylinder with free surfaces and a radius of
\SI{50}{nm} around the junction. In order to sample the triple
junction structure, we cut out a hole of radius \SI{3}{\angstrom}
around the junction center. We then randomly inserted from 8 up to 21
atoms in the hole of the donut shaped sample, using at least 10
statistically independent realizations each. For testing, we also
repeated this with structures replicated 2 or 3 times along the tilt
axis and inserted the appropriate larger amount of atoms in our
sampling.

This resulted in a range of local densities. These structures were
then minimized in two steps. First, the GB structures were made into
rigid bodies to avoid GB phase transitions and cooled from \SI{50}{K}
to \SI{0.1}{K} over \SI{50}{ps} using a Langevin thermostat with a
strong damping constant of \SI{10}{fs}. In a second step, the GBs were
made non-rigid and the system was minimized normally.

We finally tested for some triple junctions of interest that the
resulting junction was stable at room temperature. For this, we ran MD
simulations with the structure at \SI{300}{K} for \SI{1}{ns}. We
observed no structural changes.

\vspace{-0.5\baselineskip}%
\subsection{Grain boundary excess properties}
\label{sec:methods:theory}

Triple junctions cannot exist without the abutting GBs. We will thus
recapitulate the established formalisms to describe the thermodynamics
and excess properties of GBs. In the present case, the bulk phase is
fcc Al and does not undergo any phase changes. We can therefore
limit the discussion to excess properties. For GBs, their excess free
energy in a single-component system is defined as \cite{Frolov2012a}
\begin{equation}
  \label{eq:excess-GB}
  \gamma A = [U]_N -T[S]_N - \sigma_{33}[V]_N - A \sum_{i=1,2}\hat{t}_i\sigma_{3i} - \mu [N]_N.
\end{equation}
Here, $\gamma$ is the GB energy, $A$ the GB area, $U$ the internal
energy of the system, $T$ the temperature, $S$ the entropy,
$\sigma_{ij}$ the stress tensor, $V$ the volume, $\mu$ the chemical
potential, and $N$ the number of atoms in the system. The excess
shears $\hat{t}_{1,2}$ represent the translational, microscopic DOFs
of the GB and are discussed in more detail further down. We just note
here that they cannot be uniquely defined \cite{Winter2025}. The
bracket notation indicates excess properties of the GB \cite{Cahn1979,
  Frolov2012a} and is defined for any extensive property $Z$ as
\begin{equation}
  \label{eq:excess}
  [Z]_N = Z - \frac{N}{N_\text{bulk}}Z_\text{bulk}.
\end{equation}
The subscript ``bulk'' refers to a defect-free reference system.  In
this framework, it is thus $[N]_N = 0$ and the final term of
Eq.~\ref{eq:excess-GB} is zero for single-component systems.

We note that defect thermodynamics can also be expressed in the
isobaric grand canonical ensemble
\begin{equation}
  \label{eq:isobaric-grand-canonical}
  \Phi = U - TS - \sigma_{ij}V_0(\delta_{ij} + \varepsilon_{ij}) - \mu N,
\end{equation}
which is zero in equilibrium due to being fully Legendre transformed
\cite{Callen1985}. (Here, we use the infinitesimal strain
approximation with $V \approx V_0$.) For non-equilibrium defects, such
as GBs and their triple junctions, the resulting nonzero $\Phi$ is
equal to $[\Phi]$. In case the thermodynamic potential is not Legendre
transformed with regard to all stress components (e.g.,
$\varepsilon_{11}=\varepsilon_{22}=\varepsilon_{12}=0$ for our GB
definition), this equality still works if $\sigma_{ij}=0$ in the
defect-free regions for the non-transformed stresses. In this case,
only the defects contribute to the strain energy. This excess strain
energy of the defect then goes into $U$. Our definitions therefore
allow us to write
\begin{equation}
  \label{eq:excess-equals-potential}
  \gamma A = [\Phi] = \Phi.
\end{equation}
Note carefully that this is not true for other, nonzero thermodynamic
potentials, such as the Gibbs or Helmholtz free energy.

We assume that the thermodynamic (meta)stability of a triple junction
is dominated by its potential energy and simplify in the rest of this
paper by assuming $T = 0$ and $\sigma_{ij} = 0$, which can be easily
calculated with molecular statics simulations. Thus, with $U = E$,
where $E$ is the potential energy of the system, we get
\begin{align}
  \label{eq:excess-simplified}
  \Phi &\approx E - \mu N &
  &\text{and} &
  \Phi &\approx [E],
\end{align}
showing that $\mu N \approx N E_\text{bulk}/N_\text{bulk}$.

As stated in the introduction, defects can occur in different defect
phases, which can be treated with the thermodynamic framework sketched
above. For GBs, these GB phases are defined by their microscopic
DOFs. First, while $[N]_N = 0$ by definition, the GB phases can differ
by the occupancy of their crystallographic planes compared to the
defect-free crystal. While this does not affect the thermodynamics, it
leads to different GB structures and can affect the transformation
kinetics \cite{Frolov2013}.  We establish a planar fraction $\hat{n}$
for a GB of area $A$ in an orthorhombic region with $N$ atoms that has
two sides parallel to the GB. We define \cite{Frolov2013}
\begin{equation}
  \label{eq:planar-fraction}
  \hat{n} = \frac{n_\text{GB}}{n_\text{plane}}
  \quad \text{with} \quad n_\text{GB} = N~\text{mod}~n_\text{plane}.
\end{equation}
Here, $n_\text{plane}$ is the number of atoms in a defect-free
crystallographic plane with area $A$ that is parallel to the GB. Thus,
$n_\text{GB}$ represents the additional or missing atoms in the GB
plane relative to the bulk, and $\hat{n}$ expresses this as a fraction
that is independent of $A$.

\begin{figure*}
  \includegraphics[width=\linewidth]{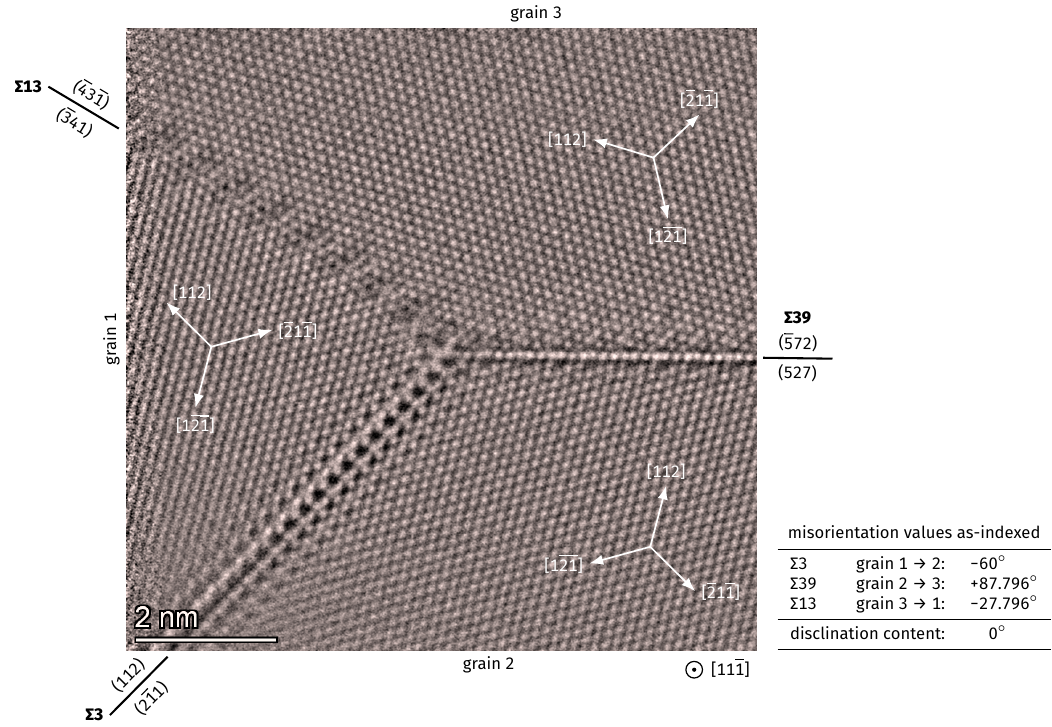}
  \caption{HAADF-STEM image of a GB triple junction in Al. The
    crystallographic $\langle112\rangle$ directions of the three
    grains were indexed and used to determine the GB planes. We found
    that the GBs are three symmetric $[11\overline{1}]$ tilt
    GBs---$\Sigma13$, $\Sigma39$, and $\Sigma3$---with the indicated
    planes. The table shows the misorientation values as indexed. Note
    that the cubic crystal has a threefold symmetry around
    $[11\overline{1}]$ and angles \ang{\pm120} are
    crystallographically equivalent. We find that the misorientations
    add up to \ang{0}, meaning that this triple junction has no
    disclination content.}
  \label{fig:exp-junc}
\end{figure*}

Additionally, the abutting crystallites can be translated against each
other. Following Winter and Frolov \cite{Winter2025}, we can define the
microscopic DOFs as
\begin{equation}
  \label{eq:translation}
  \hat{\mathbf{t}}^{(i)} = \mathbf{d}_\text{SC}^{(i)} +
  \underbrace{
    \hat{\mathbf{B}} +
    \begin{pmatrix}
      0 \\
      0 \\
      [V] + n_\text{GB} \frac{\Omega}{A} \\
    \end{pmatrix}
  }_{\hat{\mathbf{C}}}.
\end{equation}
Here, $\hat{\mathbf{B}}$ represents the crystallographic translation
between the two abutting crystallites and $\Omega$ is the equilibrium
atomic volume. In the direction normal to the GB, additional
contributions stem from the excess volume $[V]$ and from the
difference in planar occupancy in the GB. These contributions can be
summed for simplicity into $\hat{\mathbf{C}}$. There is, however, not
a unique translation vector $\hat{\mathbf{t}}$ for each GB phase: Any
shift of one of the crystallites by a displacement shift complete
(DSC) vector $\mathbf{d}_\text{SC}^{(i)}$ (which may be zero) leads to
an equivalent GB struture, albeit shifted in position. Consequently,
there are a range of equivalent, but not equal, values of
$\hat{\mathbf{t}}$. There is not an infinite number, because any
translation by a vector of the coincidence site lattice (CSL) leads to
an indistinguishable system, so $\mathbf{d}_\text{SC}^{(i)}$ is
restricted by the CSL periodicity. Furthermore, note that
$n_\text{plane} \Omega/A$ corresponds to a DSC vector, wherefore we
require $0 \leq n_\text{GB} < n_\text{plane}$ and thus
$0 \leq \hat{n} < 1$ for a unique definition of $\hat{\mathbf{C}}$.

\section{Results}
\subsection{Observed triple junction and grain boundary phases}
\label{sec:observed-junction-and-phases}

The experimentally observed GB triple junction is shown in
Fig.~\ref{fig:exp-junc}. The three adjacent grains are oriented along
a common $[11\overline{1}]$ zone axis. We indexed the
$\langle 112\rangle$ crystal directions as indicated. Then, we
generated a list of possible CSL tilt GBs with $[11\overline{1}]$ tilt
axis and $\Sigma < 100$ using the software from
Ref.~\cite{Hadian2018a}. The closest matches were the $\Sigma 3$,
$\Sigma 39$, and $\Sigma 13$ boundaries as shown in
Fig.~\ref{fig:exp-junc}. All GBs are symmetric. As-indexed, the
misorientations of these three GBs add up to \ang{0}, meaning that
there is no disclination \cite{Volterra1907, Wit1972, Romanov2003,
  King1999, Priester2013ch10} at the triple junction. This is also
evidenced by the GBs following the coincidence index combination rule
\cite{Priester2013ch10}: $3 \cdot 13 = 39$.

\begin{figure}
  \includegraphics[width=\linewidth]{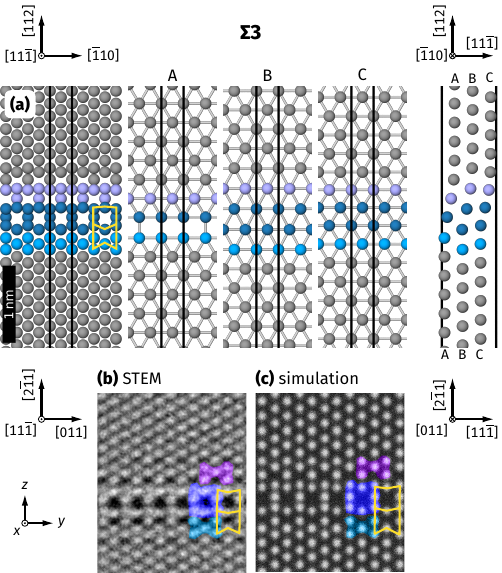}
  \caption{Atomic structure of the $\Sigma3$ $[11\overline{1}]$
    $\{112\}$ tilt GB. (a) Simulated structure. On the left, we show
    the same viewing direction as in the STEM image. The three images
    in the center show slices of the three nonequivalent
    $(11\overline{1})$ planes A, B, and C as marked in the sideview on
    the right. The gray lines in the slices represent the
    next-neighbor bonds inside the plane, highlighting that the fcc
    structure in this $\Sigma3$ GB is only disturbed on the A
    plane. Color coding of the atoms highlights the (arbitrary) atomic
    motifs, with the dark blue atoms representing the center of the
    GB.  An excerpt of the STEM image (b) is compared with a STEM
    image simulation (c). Here, we applied a stress $\tau_{31}$ to the
    latter to achieve a better match. This stress is likely to also
    occur in the real sample, see text and Supplemental
    Fig.~\ref*{fig:suppl:S3-shear}.}
  \label{fig:gb-phases:S3}
\end{figure}

In order to investigate the defect character of the junction in more
detail, we proceeded to build an atomistic computer model. First, we
ran a structure search with GRIP \cite{Chen2024} to find the three GB
structures. Detailed results of the search are listed in Supplemental
Figs.~\ref*{fig:suppl:grip-S3}--\ref*{fig:suppl:grip-S13}. All
low-energy structures have a planar fraction $\hat{n} = 0$, which
seems to be common for $[11\overline{1}]$ tilt GBs \cite{Brink2023}.

The $\Sigma 3$ $[11\overline{1}]$ $(112)$/$(2\overline{1}1)$ GB
obtained from structure search did not perfectly match the observed GB
(Supplemental Fig.~\ref*{fig:suppl:S3-shear}). Some of us reported
earlier that this $\Sigma3$ GB can have different microstates based on
the local stress state \cite{Saood2023}. This is related to the offset
between $(11\overline{1})$ planes (Fig.~\ref{fig:gb-phases:S3}(a),
right). We applied a shear stress $\tau_{31}$ and obtained the match
between STEM image and image simulation shown in
Fig.~\ref{fig:gb-phases:S3}(b,c). Such a distortion near the junction
makes sense, since the $\Sigma3$ GB is the only GB with a large offset
between $(11\overline{1})$ planes, and the two other GBs thus exert a
$\tau_{31}$ shear stress onto the $\Sigma3$ GB.

\begin{figure}
  \includegraphics[width=\linewidth]{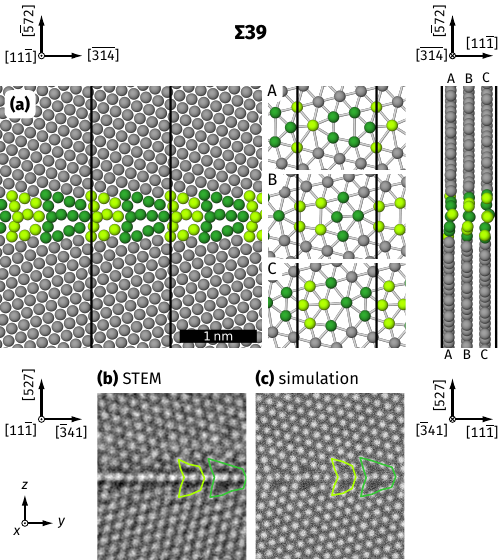}
  \caption{Atomic structure of the $\Sigma39$ $[11\overline{1}]$
    $\{257\}$ tilt GB. (a) Simulated structure. On the left, we show
    the same viewing direction as in the STEM image. The three images
    in the center show slices of the three nonequivalent
    $(11\overline{1})$ planes A, B, and C as marked in the sideview on
    the right. The gray lines in the slices represent the
    next-neighbor bonds inside the plane. Color coding of the atoms
    highlights the (arbitrary) atomic motifs.  An excerpt of the STEM
    image (b) is compared with a STEM image simulation (c).}
  \label{fig:gb-phases:S39}
\end{figure}

The $\Sigma 39$ $[11\overline{1}]$ $(\overline{5}72)$/$(527)$ GB is
clearly visible in the STEM image and matches the simulated structure
(Fig.~\ref{fig:gb-phases:S39}). We only note that the structure search
revealed another structure with \SI{3}{mJ/m^2} difference in the GB
energy at \SI{0}{K} (Supplemental
Fig.~\ref*{fig:suppl:grip-S39}). This structure, however, was not
observed experimentally. We conclude that it either represents a
microstate or that it is an artifact of the EAM potential. We use the
structure from Fig.~\ref{fig:gb-phases:S39} for the present paper.

\begin{figure}
  \includegraphics[width=\linewidth]{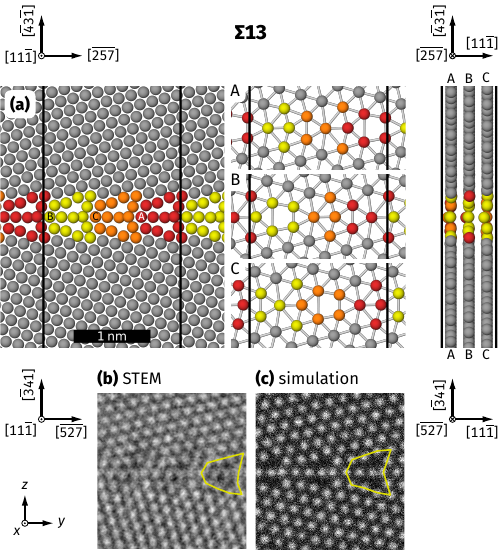}
  \caption{Atomic structure of the $\Sigma13$ $[11\overline{1}]$
    $\{134\}$ tilt GB. (a) Simulated structure. On the left, we show
    the same viewing direction as in the STEM image. The three images
    in the center show slices of the three nonequivalent
    $(11\overline{1})$ planes A, B, and C as marked in the sideview on
    the right. The gray lines in the slices represent the
    next-neighbor bonds inside the plane. Color coding of the atoms
    highlights the (arbitrary) atomic motifs.  An excerpt of the STEM
    image (b) is compared with a STEM image simulation (c).}
  \label{fig:gb-phases:S13}
\end{figure}

The $\Sigma 13$ $[11\overline{1}]$
$(\overline{4}3\overline{1})$/$(\overline{3}41)$ GB exhibits some
moir\'e patterns in the STEM image along the GB, which likely indicate
that there is a GB step in the depth of the sample.
We nevertheless compared the minimum-energy structure obtained in the
structure search with the experimental results in
Fig.~\ref{fig:gb-phases:S13}. The motif that we marked in yellow in
Fig.~\ref{fig:gb-phases:S13}(b) is the same as in the STEM image
simulation of the model structure, even though there is some noise in
the center of the motif in the experiment. We further note from the
simulation results that this motif is repeated three times in the
periodic unit cell of the GB, being shifted each time by one atomic
layer along the $[11\overline{1}]$ axis.

These GB motifs were observed and characterized earlier
\cite{Saood2024} and are typical for $[11\overline{1}]$ tilt GBs in
Al. There, motifs with no offset between $(11\overline{1})$ planes
were classified as ``bow \& arrow'' structures, the others as
``zipper''. All these GB phases have in common that
$\hat{n} = n_\text{GB} = 0$ and thus $\hat{C}_{1,2} = \hat{B}_{1,2}$;
$\hat{C}_3 = \hat{B}_3 + [V]$ (see Eq.~\ref{eq:translation}).

\subsection{Computer model of the junction and its energy}
\label{sec:computer-model-of-junction}

\begin{figure}[t]
  \centering
  \includegraphics[width=\linewidth]{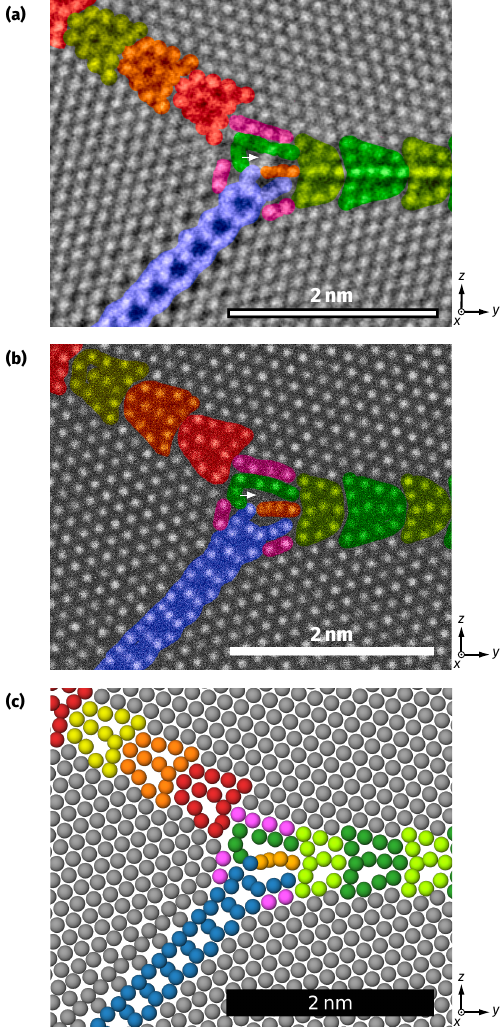}
  \caption{Comparison of the experimental junction with the computer
    model. (a) Excerpt of the experimental STEM image, with false
    colors to highlight motifs. (b) STEM image simulation of the
    computer model at \SI{300}{K} with the same false color
    highlights. (c) Snapshot of the model structure at \SI{0}{K}.}
  \label{fig:compare-junc-struct}
\end{figure}

To construct a triple junction, we only need to rotate the obtained GB
structures and join them. While there is no disclination, and
therefore no gap due to angular mismatch, it stands to reason that the
crystallites still cannot be joined perfectly. This is because the
crystallites at a GB can be translated with regard to each other
(microscopic DOFs).

In a first step, we thus tried to reproduce the motifs of the
experimental junction. We marked certain motifs around the junction
area in the STEM image (Fig.~\ref{fig:compare-junc-struct}(a), pink
atomic columns).  The simulated GB structures were then shifted,
combined, and overlapping atoms deleted to reproduce the arrangement
of atomic motifs next to the junction core. Afterwards, a cylindrical
hole was cut around the junction. For sampling, this hole was then
randomly filled with various amounts of atoms and minimized. Our best
match to the experimental structure is shown in
Fig.~\ref{fig:compare-junc-struct}. We performed STEM image
simulations on two snapshots with $L_x = \SI{21.1}{nm}$ (30 unit cells
along the tilt axis) that were equilibrated at \SI{300}{K} to capture
the effect of phonons and thermal expansion. The STEM image simulation
matches almost perfectly to the experimental image, except for the
increased intensity in the center of the junction (arrows). The latter
is likely due to a step of the $\Sigma13$ GB along the depth of the
experimental sample (as discussed before in
Sec.~\ref{sec:observed-junction-and-phases}).

\begin{figure}
  \includegraphics[width=\linewidth]{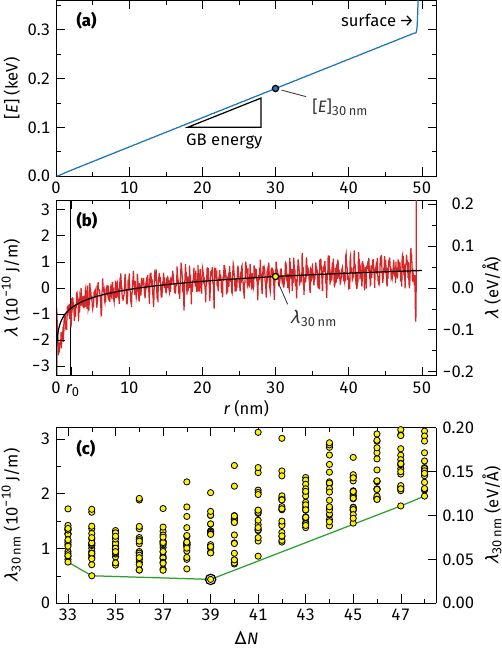}
  \caption{(a) Cumulative excess energy $[E]$ inside the given radius
    $r$. The slope is approximately equal to the sum of GB energies
    (see text). For comparison between different structures, we use
    the excess energy at $r = \SI{30}{nm}$ (data point).  (b) The line
    energy $\lambda$ follows a logarithmic trend. (c) Structure
    sampling with different numbers of atoms $\Delta N$ inserted into
    the triple junction region. These simulations were performed for a
    sample with a thickness of 3 unit cells along the $x$
    direction. Panels (a) and (b) show the data for the marked
    structure with the lowest line energy at $\Delta N = 39$.}
  \label{fig:line-energy-calc}
\end{figure}

\begin{figure*}
  \includegraphics[width=\linewidth]{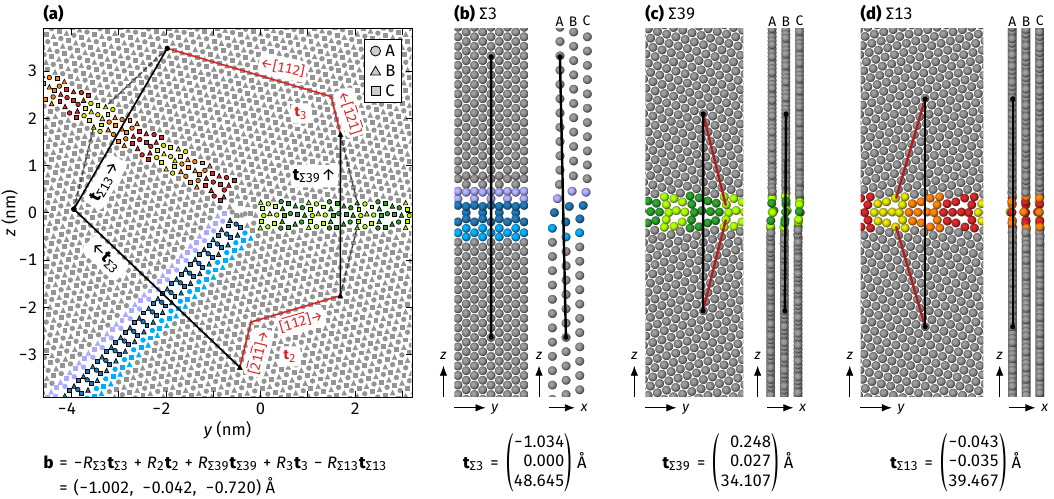}%
  \vspace{0.5\baselineskip}   
  \caption{(a) Burgers circuit around the triple junction that fits
    best to the experiment (Fig.~\ref{fig:compare-junc-struct}). The
    circuit goes counter-clockwise around the $x$ direction
    ($[11\overline{1}]$ tilt axis). The red segments $\mathbf{t}_2$ and
    $\mathbf{t}_3$ of the circuit move through defect-free fcc
    regions. The black segments $\mathbf{t}_{\Sigma3}$,
    $\mathbf{t}_{\Sigma39}$, and $\mathbf{t}_{\Sigma13}$ must be
    measured from defect-free reference GB structures in (b)--(d).
  }
  \label{fig:burg-circ}
\end{figure*}

In order to rationally select a ``best'' structure from our sampling,
we need to define a junction energy. We stay in the zero Kelvin
approximation for simplicity and assume unstrained bulk
crystallites. As for GBs, we can use the excess energy
$[E] = E - \mu N$ (see
Eqs.~\ref{eq:excess}--\ref{eq:excess-simplified}). Here, $E$ is the
total potential energy of the sample, $N$ is the total number of atoms
in the sample, and $\mu$ the chemical potential, which is equal to the
bulk's cohesive energy per atom at zero temperature and stress. This
excess energy is calculated in a cylindrical region around the triple
junction. On the atomic scale, however, the triple junction is not a
line, but has a certain thickness. To be able to compare different
triple junction core structures, we chose a uniform definition of the
center line around which we draw the cylinder. We use the
intersections of the GB planes as demonstrated in Supplemental
Fig.~\ref*{fig:suppl:energy-dependence-on-center}(a).

Figure~\ref{fig:line-energy-calc}(a) shows the excess energy
calculated within a cylindrical region of radius $r$ around the
assumed junction line.  We can see that $[E](r)$ has an approximately
linear slope, which is close to the combined GB excess energy
$(\gamma_{\Sigma3} + \gamma_{\Sigma39} + \gamma_{\Sigma13}) L_x$.

To extract the line energy, we define the GB area to extend from the
triple junction to the radius $r$ for each GB. For a thickness $L_x$
of the sample along the tilt axis, we can thus define a line energy
\begin{equation}
  \label{eq:line-energy}
  \lambda = \frac{[E] - (\gamma_{\Sigma3} + \gamma_{\Sigma39} + \gamma_{\Sigma13}) A}
                 {L_x}.
\end{equation}
(Note that we always exclude the region close to the surface to avoid
an influence of the surface energy.) This is shown in
Fig.~\ref{fig:line-energy-calc}(b). We see that the triple junction
energy is not constant, but depends on the radius, indicating that the
triple junction is connected to a long-range strain field that we will
discuss further down.

For comparability, we used the line energy $\lambda_{\SI{30}{nm}}$ at
radius \SI{30}{nm} to compare the energies of different structures
during sampling (Fig.~\ref{fig:line-energy-calc}(c)). We chose this
radius since it is sufficiently far away from the surface
($r = \SI{50}{nm}$) and the junction center. The lowest-energy
structure corresponds to the best fit with experiment from
Fig.~\ref{fig:compare-junc-struct}. We verified that a different
choice of junction line center shifts all line energies by the same
amount (Supplemental
Fig.~\ref*{fig:suppl:energy-dependence-on-center}), indicating that
this introduces a systematic error and virtually no scatter, meaning
that the choice of minimum-energy structure is robust with regards to
choice of the center. Furthermore, there is a direct, linear
correlation between $[E]_{\SI{30}{nm}}$ and $\lambda_{\SI{30}{nm}}$ so
that the minimum-energy structure with either criterion is the same
(Supplemental Fig.~\ref*{fig:suppl:relation-excess-to-line-energy}).
Additionally, we repeated the structure search with one, two, and
three unit cells along the periodic $x$ direction, yielding the same
results (Supplemental
Fig.~\ref*{fig:suppl:junction-search-thicknesses}).

\subsection{Dislocation character of the triple junction}
\label{sec:dislocation-character}

The logarithmic radial dependence of the line energy
(Fig.~\ref{fig:line-energy-calc}(b)) implies a dislocation-like nature
of the triple junction. We therefore define a Burgers circuit to
extract the defect content of the junction line over the pure GBs. The
dislocation content of the individual GBs depends on the macroscopic
misorientation (see, e.g., Read and Shockley \cite{Read1950}), but
does not lead to long-range strain fields \cite{Nabarro1952}. To
separate the junction Burgers vector content from the GBs, we will
therefore make a circuit with reference to defect-free GBs. This was
proposed for GB phase junctions before \cite{Frolov2021}, but can be
trivially extended to triple junctions.

For this, we first construct a closed loop over the triple junction in
Fig.~\ref{fig:burg-circ}(a). The red lines $\mathbf{t}_2$ and
$\mathbf{t}_3$ move only through fcc regions and can be calculated the
same way as for a classical Burgers circuit. The GB crossings are
measured in otherwise defect-free bicrystals
(Fig.~\ref{fig:burg-circ}(b)--(d)). All line segments $\mathbf{t}_i$
are then rotated into the simulation's coordinate system as depicted
in Fig.~\ref{fig:burg-circ}(a) using the relevant rotation matrices
$R_i$. We obtain
\begin{equation}
  \label{eq:burgers-observed}
  \begin{split}
  \mathbf{b}_\text{obs}
             = &-R_{\Sigma3}\mathbf{t}_{\Sigma3} \\
               &+R_2 \left(6\frac{a}{6}[\overline{2}1\overline{1}]
                           +12\frac{a}{6}[\overline{112}] \right) \\
               &+R_{\Sigma39}\mathbf{t}_{\Sigma39} \\
               &+R_3 \left(5\frac{a}{6}[\overline{1}2\overline{1}]
                           +22\frac{a}{6}[112]
                           -\frac{a}{3}[11\overline{1}] \right) \\
               &-R_{\Sigma13}\mathbf{t}_{\Sigma13}
               = (-1.002, -0.042, -0.720)\,\text{\AA}.
  \end{split}
\end{equation}
We checked that this Burgers vector is independent of the structure
and amount of atoms inserted into the junction region. We will explore
the relation of the junction's defect content to the GB DOFs
later. Note that the GB crossings require very exact length
measurements to obtain the correct Burgers vector, which is
unfortunately not possible in the experiment given the measurement
uncertainties and possible sample distortions.

Due to the junction line's dislocation character, the energy depends
on the cylinder radius as follows \cite{HirthLothe1992}:
\begin{equation}
  \label{eq:disloc-energy}
  \lambda = Kb^2 \ln\left(\frac{r}{r_0}\right) + \lambda_0.
\end{equation}
Here, $K$ is a prefactor including the elastic moduli of the bulk,
$r_0 = \SI{2}{nm}$ is the (arbitrary) core radius, and $\lambda_0$ is
the core energy. We chose the core radius to include all disturbed
structural GB motifs near the junction. We obtain $K=\SI{3.0}{GPa}$
and $\lambda_0= \SI{-0.79e-10}{J/m}$.

The meaning of negative line energies (or core energies in the present
case) has been debated. As already pointed out by King
\cite{King1999}, the triple junction can easily be energetically
favorable compared to the GBs, as long as it is still of higher energy
than the defect-free bulk region. A negative line energy in no way
means that the system is unstable against the formation of more triple
junctions, since the formation of a triple junction necessarily
requires the formation of the abutting GBs. In other words, the
separation of $[E]$ into GB energies $\gamma$ and a line energy
$\lambda$ is arbitrary, but useful to understand the dislocation-like
behavior of the junction. While the absolute defect energy of the
junction line is therefore ill-defined, energy differences between
junctions are nevertheless meaningful, as we saw before in
Supplemental Figs.~\ref*{fig:suppl:energy-dependence-on-center} and
\ref*{fig:suppl:relation-excess-to-line-energy}. Importantly, due to
the logarithmic dependence in Eq.~\ref{eq:disloc-energy}, the line
energy diverges to positive infinity with increasing system size.  The
line energies are consequently always positive for sufficiently large
system sizes (or grain sizes in a polycrystal).

Finally, the interatomic potential has a shear modulus
$\SI{17.6}{GPa} \leq G \leq \SI{44.1}{GPa}$ (range of elastic
anisotropy) and a Poisson ratio $\nu = 0.33$. Assuming a screw
dislocation with $K = G / (4\pi)$ \cite{HirthLothe1992} we estimate
$G = \SI{37.3}{GPa}$ from Eq.~\ref{eq:disloc-energy}, while we obtain
$G = \SI{24.8}{GPa}$ for an edge dislocation ($K = G / (4\pi(1-\nu))$
\cite{HirthLothe1992}). This lies in the expected range and further
confirms the dislocation character of the GB triple junction.

\begin{figure*}
  \includegraphics[width=\linewidth]{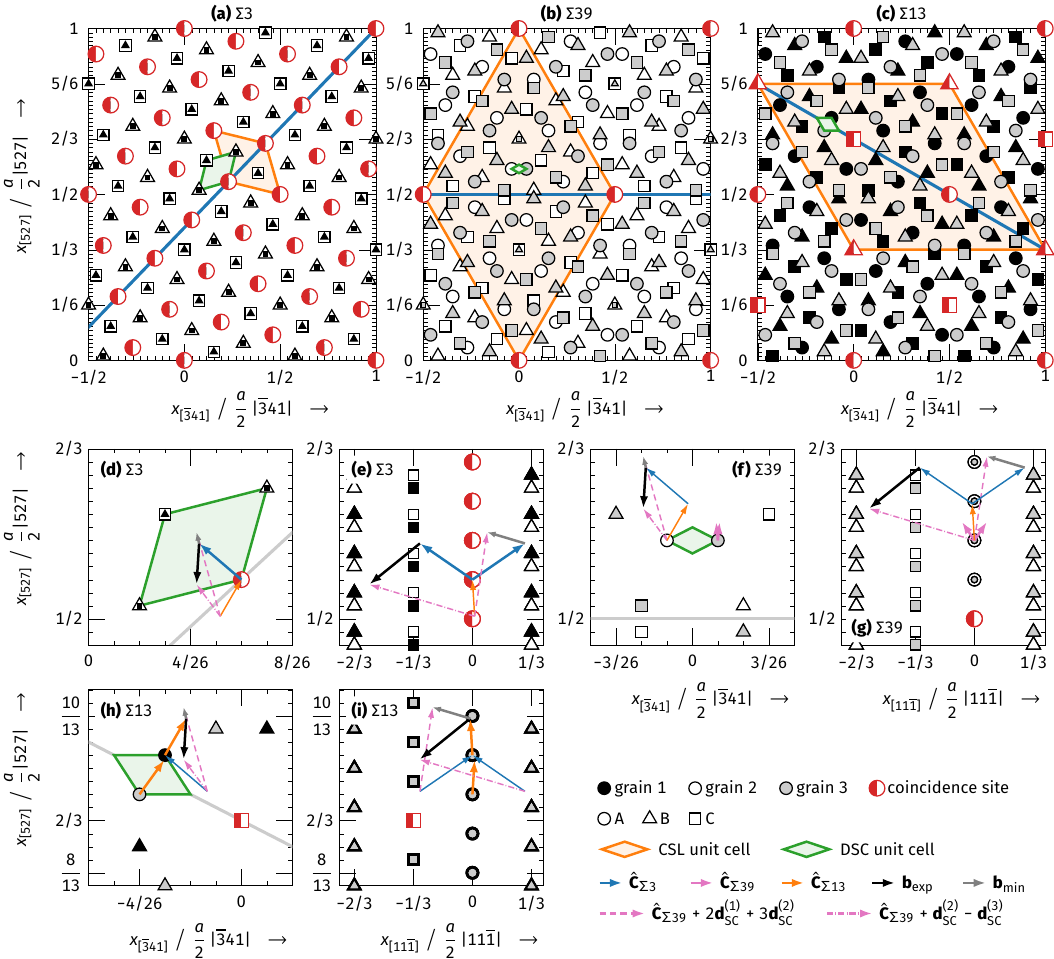}
  \caption{Dichromatic patterns for the GBs. The indicated crystal
    directions belong to grain 2, i.e., the bottom crystal of the
    $\Sigma39$ GB. (a)--(c) Full dichromatic patterns, with CSL and
    DSC unit cells indicated. (d), (f), (h) Zooms on the DSC unit
    cells. (e), (g), (i) Sideviews from the $-[\overline{3}41]$
    direction.  In panels (d)--(i), the smallest possible Burgers
    vector and the experimentally observed Burgers vector are drawn
    onto the dichromatic patterns in gray and black. They are shown to
    be sums of the microscopic DOFs $\hat{\mathbf{t}}$ of the three
    GBs (colored arrows). In our chosen representation, $\Sigma3$ and
    $\Sigma13$ have $\hat{\mathbf{t}} = \hat{\mathbf{C}}$, while the
    $\Sigma39$ GB's DOFs contain additional DSC vectors (dashed
    arrows).}
  \label{fig:dichrom}
\end{figure*}

\subsection{Sampling the microscopic degrees of freedom}

We already noted that the Burgers vector is independent of the triple
junction structure. It must therefore be a result of the GBs' DOFs.
While each GB structure (GB phase) is connected to specific DOFs
$\hat{\mathbf{C}}$ (see Sec.~\ref{sec:methods:theory}), there are many
crystallographically equivalent realizations $\hat{\mathbf{t}}$ of the
same GB structure. The reason is that shifting one crystallite by a
DSC vector will result in the same atomic arrangement within the GB,
albeit potentially on a different parallel plane in the crystal. This
is limited by the CSL. Shifting one crystallite by exactly a CSL
vector will leave behind a completely undistinguishable atomic
structure (assuming an infinite medium, otherwise there would of
course be a surface step). Thus, each GB has a large, but finite
amount of DOFs that affect how the three crystallites fit together at
the triple junction. If a gap remains, a Burgers vector will result
because the crystallites have to deform with a long-range elastic
field to close the misfit. Assuming that we express our microscopic
DOFs $\hat{\mathbf{t}}$ as the displacement of the upper crystallite
compared to the lower crystallite, and assuming a counter-clockwise
Burgers circuit around the junction as depicted in
Fig.~\ref{fig:exp-junc}, we can express the resulting Burgers vector
as
\begin{equation}
  \label{eq:burgers-vector}
  \mathbf{b} = -R_{\Sigma3}  \hat{\mathbf{t}}_{\Sigma3}
               +R_{\Sigma39} \hat{\mathbf{t}}_{\Sigma39}
               -R_{\Sigma13} \hat{\mathbf{t}}_{\Sigma13}.
\end{equation}
Here, $R_i$ are the rotation matrices to align the GBs as found in the
experimental triple junction. This is similar to the derivation of the
Burgers vector at a GB phase junction \cite{Frolov2021}, where two
different GB structures/chemistries meet, albeit with three joining
GBs.

For the given GBs, the CSL and DSC unit cells are shown in
Fig.~\ref{fig:dichrom}(a)--(c) and their basis vectors listed in
Supplemental Table~\ref*{tab:suppl:csl-dsc}. We can thus explore
different combinations of the microscopic DOFs
$\hat{\mathbf{t}} = \hat{\mathbf{C}} + \mathbf{d}_\text{SC}$, listed
in Supplemental Tables~\ref*{tab:suppl:csl-dsc} and
\ref*{tab:suppl:dof}, to obtain different Burgers vectors. The model
of the experimentally-observed junction has a Burgers vector of
\begin{equation}
  \label{eq:exp-burgers}
  \begin{split}
    \mathbf{b}_\text{exp} = &-R_{\Sigma3}  \hat{\mathbf{C}}_{\Sigma3^-} \\
                            &+R_{\Sigma39} \left(
                              \hat{\mathbf{C}}_{\Sigma39^+}
                              +\mathbf{d}_\text{SC}^{\Sigma39,(2)}
                              -\mathbf{d}_\text{SC}^{\Sigma39,(3)}
                            \right)\\
                            &-R_{\Sigma13}\hat{\mathbf{C}}_{\Sigma13^-} \\
                          = &-R_{\Sigma3}  \hat{\mathbf{C}}_{\Sigma3^-}
                            +R_{\Sigma39}\hat{\mathbf{C}}_{\Sigma39^+} \\
                            &-R_{\Sigma13} \left(
                              \hat{\mathbf{C}}_{\Sigma13^-}
                              +\mathbf{d}_\text{SC}^{\Sigma13,(3)}
                            \right)\\
                          = & \,(-1.002, -0.040, -0.717)\,\text{\AA}
                          \approx \mathbf{b}_\text{obs}.
  \end{split}
\end{equation}
The experimentally observed triple junction thus has mixed edge/screw
character, with comparable components along the triple junction line
($x$) and normal to it ($z$).  We can see that there are different
possible combinations of $\hat{\mathbf{t}}$ for the three GBs that sum
to the same Bur\-gers vector. Here, we show two. As expected, the result
matches the measured Burgers vector $\mathbf{b}_\text{obs}$
(Eq.~\ref{eq:burgers-observed}). Small differences are due to
uncertainties when measuring the GB crossings for the Burgers circuit
(Fig.~\ref{fig:burg-circ}(b)--(d)).

The minimal possible Burgers vector is
\begin{equation}
  \label{eq:min-burgers}
  \begin{split}
    \mathbf{b}_\text{min} = &-R_{\Sigma3}  \hat{\mathbf{C}}_{\Sigma3^+} \\
                            &+R_{\Sigma39} \left(
                              \hat{\mathbf{C}}_{\Sigma39^+}
                              +2\mathbf{d}_\text{SC}^{\Sigma39,(1)}
                              +3\mathbf{d}_\text{SC}^{\Sigma39,(2)}
                            \right)\\
                            &-R_{\Sigma13}\hat{\mathbf{C}}_{\Sigma13^-} \\
                            = & \,(-0.743, -0.040, 0.196)\,\text{\AA}.
  \end{split}
\end{equation}
It has its largest component along the tilt axis, which is also the
junction line direction, and has thus a large screw component. Both
$\mathbf{b}_\text{exp}$ and $\mathbf{b}_\text{min}$ are illustrated in
Fig.~\ref{fig:dichrom}(d)--(i).

There should also be a maximum Burgers vector due to the CSL
periodicity, i.e., $\hat{\mathbf{t}}$ gets wrapped back into the CSL
unit cell. When iterating over the non-equivalent $\hat{\mathbf{t}}$
and computing the resulting Burgers vectors, we still obtain some
Burgers vectors that are so large that they exceed the periodicity of
the fcc lattice. Consequently, the largest possible triple junction
Burgers vector is on the order of the fcc bond length, which is equal
to the length of the primitive unit cell vectors of fcc. Thus
$|\mathbf{b}| \leq \SI{2.85}{\angstrom}$ in Al, which corresponds to a
full $a/2 \langle011\rangle$ bulk dislocation.

\begin{figure}
  \includegraphics[]{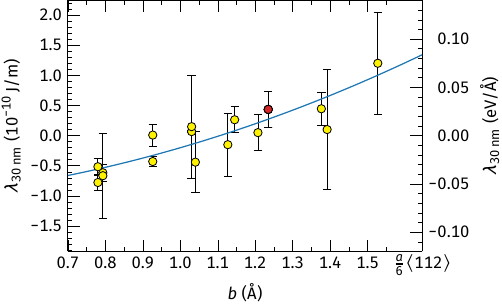}
  \caption{Smallest line energies at $r = \SI{30}{nm}$ as found by
    structure search for a range of Burgers vectors. The red data
    point belongs to the experimentally observed triple junction
    discussed in Sec.~\ref{sec:computer-model-of-junction}. The error
    bars indicate the range of the systematic error due to different
    choices of the triple junction center line (see Supplemental
    Fig.~\ref*{fig:suppl:energy-dependence-on-center}). The blue line
    is a fit of a parabola to the data.}
  \label{fig:line-energy}
\end{figure}

We move on to build several computer models of junctions with
different Burgers vectors. The predicted and measured Burgers vectors
matched in all cases. The resulting line energies are plotted in
Fig.~\ref{fig:line-energy} as a function of the magnitude of the
Burgers vector. While there is scatter due to the differences in core
energies, the trend is roughly $\lambda \propto b^2$, as expected for
dislocations. As already discussed in
Sec.~\ref{sec:dislocation-character}, negative line energies can
occur, based on the system size for which the line energy is
calculated. This simply means that the localized region around the
junction has a lower excess energy than the joining GBs.

Using Eq.~\ref{eq:disloc-energy}, we can again verify if
the prefactor $K$ fits to the data with $K = G / (4\pi)$ for screw
dislocations \cite{HirthLothe1992} and $K = G / (4\pi(1-\nu))$ for
edge dislocations \cite{HirthLothe1992}. The interatomic potential has
a shear modulus $\SI{17.6}{GPa} \leq G \leq \SI{44.1}{GPa}$ (range of
elastic anisotropy) and a Poisson ratio $\nu = 0.33$. Assuming a screw
dislocation, the fit in Fig.~\ref{fig:line-energy} corresponds to a
shear modulus of \SI{42}{GPa}, while assuming an edge dislocation
corresponds to $G = \SI{28}{GPa}$. This lies in the expected range and
further confirms the dislocation character of the GB triple junction.

These results demonstrate that the triple junction has a dislocation
character that is determined by the microscopic DOFs of the three
joining GBs. The triple junction line energy follows the same trends
as the energy of a bulk dislocation.

\subsection{Comparison to triple junction energies in literature}

How do our line energies $\lambda$ compare to previous results in the
literature?  Measurements of surface grooves at the triple junction
\cite{Gottstein2010} yielded values on the order of \SI{6e-9}{J/m} for
Cu. This number is an order of magnitude higher than the range that we
found (Figs.~\ref{fig:line-energy-calc} and
\ref{fig:line-energy}). However, it is unclear to what region around
the groove the measurement is sensitive, and part of what we assigned
to the GB energy may be contained in this number. Computer simulations
in fcc metals \cite{Eich2016, Tuchinda2024} found values on the order
of \SI{-e-10}{J/m}, more in line with our results. However, depending
on the Burgers vector and the radius $r$ within which we measured the
line energy including the elastic strain energy, we obtain either
negative or positive values.

The literature values cannot be directly compared to ours, because
these authors assumed a constant line energy without long-range
elastic fields in their calculations. This is clearly visible, e.g.,
in Fig.~3 of Ref.~\cite{Eich2016}, where the elastic strain energy was
assigned to the GBs by assuming that the GB energy converges with
$r$. In the finite-size simulation cell, it indeed looks as if the GB
energy defined this way converges, but due to the logarithmic function
of the strain energy, it actually diverges in infinitely-sized
systems.

\section{Conclusion}

We observed a GB triple junction in a $\{ 111\}$-textured Al thin film
with atomic resolution. We analyzed the defect character of the triple
junction in more detail using a surrogate atomistic computer
model. Our findings are:

\begin{itemize}
\item Triple junctions have dislocation character, i.e., a Burgers
  vector. They are therefore connected with long-range strain
  fields. We presented a method to measure the Burgers vector directly
  using GB reference structures. Additionally, we demonstrated for our
  example of a disclination-free triple junction that this Burgers
  vector is a result of the microscopic DOFs of the joining GBs. There
  are a large number of possible combinations for the same three GBs,
  but the number is not infinite. This is because the periodicity of
  the CSL limits the distinguishable combinations.
\item The magnitude of the Burgers vector is comparable to that of
  disconnections and not much smaller than for bulk dislocations
  ($b_\text{min}$ is half as big as a Shockley partial). On the one
  hand, this indicates that the triple junction can act as a sink or
  source for other line defects, and on the other hand, elastic
  interactions between bulk dislocations, disconnections, and triple
  junctions will play an important role in GB network evolution and
  plasticity.
\item A direct consequence of the triple junction's strain field is
  the divergence of its line energy for infinitely-sized
  systems. Using an EAM potential on the computer models, we showed
  that the line energy follows the same logarithmic law as bulk
  dislocations. Any line energy that was reported as a single value in
  the literature is thus only meaningful when the corresponding system
  size is known. For different realizations of our triple junction, as
  well as for different system sizes, we observed both positive and
  negative line energies. Within a radius $r \leq \SI{50}{nm}$ around
  the junction, the line energies were on the order of \SI{\pm
    e-10}{J/m}.
\item Interestingly, the experimentally observed triple junction does
  not have the lowest possible Burgers vector and thus not the lowest
  possible line energy. It stands to reason that a transformation of
  the triple junction requires coordinated movement of the GBs and
  potentially diffusion to rearrange the triple junction core
  region. The kinetics of this process are likely too slow to be
  driven by the small energy contribution of the triple junction
  within the experimental time frame. Additionally, any triple
  junction is part of a GB network. Shifting the microscopic DOFs of
  one GB will thus affect \emph{two} triple junctions, coupling their
  excess energies. In combination with elastic junction interactions,
  this interdependence means that any energy minimization affects the
  whole GB network. Consequently, we expect that the junction types in
  complex GB networks are the result of junction interactions and
  sample history, and are thus not accurately predictable.
\end{itemize}

\section{Data availability}

Data to be published before final publication.\\

\section{Acknowledgments}

Part of this work was supported by the Deutsche Forschungsgemeinschaft
(DFG) within the SFB 1394, ``Structural and Chemical Atomic
Complexity: From Defect Phase Diagrams to Material Properties''
(project ID 409476157).%

\textit{Author contributions:} TB conducted the computer simulations,
analyzed the data, and wrote the initial manuscript draft. SS
performed the STEM experiments. PS assisted in the analysis of the
STEM results and implemented the STEM image simulations. TB, JN, and
GD conceptualized the research. JN and GD secured funding within the
SFB 1394, supervised the project, and contributed to discussions at
all stages. All authors contributed to the preparation of the final
manuscript.

\end{document}